\documentclass[pre,twocolumn,preprintnumbers,amsmath,amssymb]{revtex4-1}

\usepackage{amsfonts}
\usepackage{bbold}
\usepackage{mathrsfs}
\usepackage{color}
\usepackage{epstopdf}

\usepackage{graphicx}
\usepackage{bm}

\usepackage{amssymb, dsfont}

\newcommand{\mttp}[1]{\textcolor{black}{#1}}

\newcommand{\nn}{\nonumber }

\newcommand{\rr}{{\mathbf r}}
\newcommand{\kk}{{\mathbf k}}

\newcommand{\p}{\partial}

\newcommand{\BEQ}{\begin{equation}}
\newcommand{\EEQ}{\end{equation}}
\newcommand{\BEA}{\begin{eqnarray}}
\newcommand{\EEA}{\end{eqnarray}}

\begin{document}




\title{\mttp{Effective equilibrium picture in $xy-$model with exponentially correlated noise}}

%

\author{Matteo Paoluzzi$^{1}$}
\email{mpaoluzz@syr.edu}                                  
\author{Umberto Marini Bettolo Marconi$^{2}$}
\author{Claudio Maggi$^{3}$}

\affiliation{
$^1$ Department of Physics and Syracuse Soft Matter Program, Syracuse University, Syracuse NY 13244, USA  \\
$^2$ Scuola di Scienze e Tecnologie, Universit\`a di Camerino, Via Madonna delle Carceri, 62032, Camerino, INFN Perugia, Italy \\
$^3$ NANOTEC-CNR, Institute of Nanotechnology, Soft and Living Matter Laboratory, Piazzale A. Moro 2, I-00185, Roma, Italy  
}

\begin{abstract}
We study the effect of exponentially correlated noise on $xy-$model 
in the limit of small correlation time discussing the order-disorder transition in mean field
and the topological transition in two dimensions.
We map the steady states of the non-equilibrium dynamics into an effective equilibrium theory. 
In mean-field, the critical temperature increases with the noise correlation time $\tau$
indicating that memory effects promote ordering. This finding is confirmed by numerical simulations.
The topological transition temperature in two dimensions remains untouched. However, finite size 
effects induce a crossover in the vortices proliferation that is confirmed by numerical simulations.  
\end{abstract}
\maketitle

\section{Introduction}

\mttp{The classical $xy-$model undergoes a phase transition that is second order in $d>2$ spatial 
dimensions and infinite order in $d=2$ \cite{Plischke,Ma}. 
Since $xy-$model is described by a vectorial order parameter 
invariant under $O(2)$ orthogonal symmetry group, the existence
of a second order phase transition in $d$ spatial dimensions is 
governed by Mermin-Wagner theorem that fixes the lower critical
dimension at $d=2$ \cite{MerminWagner,Coleman,Hohenberg}.
However, in two dimensions, topological defects produce collective configurations
like vortices that cause a novel type
of phase transition related to the vortex/anti-vortex pair unbinding \cite{B1,B2,KT}, 
i. e., the so-called Berezinskii-Kosterlitz-Thouless transition (BTK).
%
%
}

\mttp{
In this paper, we investigate the properties of $xy-$model driven out-of-equilibrium through 
exponentially correlated noise. 
The control parameters of the dynamics are the correlation time of the noise $\tau$ and the
strength of the noise $T$. When $\tau=0$, the model reduces to the equilibrium $xy-$model
at temperature $T$.
By considering Unified Colored Noise Approximation (UCNA) \cite{Jung87,Maggi} in the 
small $\tau$ limit, we write
an effective equilibrium theory that is exact in the small $\tau$ limit. 
In the effective equilibrium picture, $\tau$ becomes an external thermodynamic parameter that
can be tuned to bring the system to the transition point.
}

\mttp{ 
%
We will start by discussing the model in mean-field approximation corresponding
to $d=\infty$. To do so, we consider a fully-connected lattice \cite{Parisi_stat}.
%
In the small $\tau$ limit, we can compute analytically the partition function obtaining
a vectorial $O(2)$ field-theory where the Landau parameters depend on both, temperature and $\tau$.  
According to that finding, the mean-field model for small $\tau$ undergoes a second order phase transition
at a $\tau$ dependent temperature. 
We show that, as well the scalar field theories \cite{Paoluzzi},
exponentially correlated noise promotes order in the
sense that the resulting mean-field critical temperature $T_{mf}(\tau)$ is an increasing
function of $\tau$, i. e., by increasing $\tau$ the critical temperature increases too. 
}

\mttp{
After that, we will address the problem in $d=2$, where for $\tau=0$ the second order phase transition is
replaced by BTK transition at temperature $T_{BKT}$.  
In that case, the effective equilibrium picture is obtained 
considering the continuum limit of $xy-$model, i. e., in the spin-wave approximation.
From the computation of the 
spatial correlation function,
we show that no long-range order can be obtained
at small but finite $\tau$.
The impact of $\tau$ on BKT will be investigated considering the single vortex energy cost. 
Even though the effect of correlated noise becomes negligible in the thermodynamic 
limit, we find a linear shift at higher temperature $T_{BKT}$ that scales logarithmically 
in the system size.
}

\mttp{
In Active Matter \cite{Bechinger17,Marchetti13,Vicsek12,cates2012diffusive,Reyes}, recent works pointed out the importance of memory
effects on the angular dynamics of Vicsek like models \cite{Nagai15,Sumino12}.
However, in the presence of memory effects, it is not possible to perform the
usual coarse graining procedure to obtain hydrodynamic equations \cite{Nagai15,Marchetti13}.
In the model we are going to consider, since the calculation is performed on a lattice, 
density fluctuations are not taken into account. However, the 
effective equilibrium picture could be extended to off-lattice model.}

\mttp{We also perform numerical simulations to check the validity of the approximated solution.}
We compare the predictions given by the approximated theory 
with numerical simulations for both cases, mean-field
and two dimensions. In particular, the theoretical expression for the critical temperature
in mean-field is in good agreement with numerical simulations. In two dimensions,
we recover a linear shift in $\tau$, in agreement with the prediction of the theory.

\section{The Model}
We consider the dynamics of a two 
dimensional $xy-$model driven by exponentially
correlated noise. The system is composed by $N$
compasses $\mathbf{s}_i=(\cos\theta_i,\sin\theta_i)$,
with $i=1,...,N$, arranged on a two dimensional square lattice. 
The model can be introduced formally by considering the following 
equation of motion for the angular degree of freedom $\theta_i$
\BEQ \label{gcn}
\dot{\theta}_i  = -\frac{\p H_{xy}}{ \p \theta_i} + \zeta_i \, .
\EEQ
The Hamiltonian is
\BEQ \label{xy}
H_{xy}[\theta] = -\frac{1}{2}\sum_{i,j} J_{ij} \cos(\theta_i - \theta_j)
\EEQ 
where $J_{ij}$ is the adjacency matrix. In mean-field, $J_{ij}=J/N,$ $\forall i,j=1,..,N$, i .e., fully-connected lattice. 
In $d$ dimensions, $J_{ij}=J$ is different from zero only for nearest neighbors sites. 
We consider ferromagnetic coupling
$J>0$. 
%
The noise term $\zeta_i$ is colored and Gaussian  
\BEA
&&\langle \zeta_i (t) \rangle=0 \\ \nn
&&\langle \zeta_i(t) \zeta_j(s) \rangle=\frac{2 T}{\tau} \delta_{ij} e^{- \frac{|t - s|}{\tau} }
\EEA

To start our analytical computation we rewrite (\ref{gcn}) using
an auxiliary variable $\psi_i$ for each angular degree of freedom $\theta_i$.
To ensure an exponentially correlated dynamics for $\theta_i$, $\psi_i$ 
undergoes an Ornstein-Uhlenbeck process. 
We can recast the original equations of motion (\ref{gcn}) as follows
\BEA \label{motion}
\dot{ \theta}_i  &=& -\frac{\p H_{xy}}{\p \theta_i} + \psi_i \, , \\ \nn
\tau \, \dot{\psi}_i  &=& -\psi_i + \sqrt{T} \eta_i \, .
\EEA
Now the noise term $\eta_i$ is white and Gaussian, i. e., $\langle \eta_i \rangle = 0$ and $\langle \eta_i(t) \eta_j (s) \rangle=2 \delta_{ij} \delta(t-s)$. 
$T$ tunes the strength of the noise, $\tau$ is the persistence time.
When $\tau=0$, our model reduces to the equilibrium $xy-$model at temperature $T$. 
In the opposite limit, i. e., $\tau\to\infty$ \mttp{and $T$ finite}, $\psi_i$ is a random and quenched variable
and we recover the Kuramoto model \cite{Kuramoto_review}.
\mttp{It is wort noting that Eqs. (\ref{motion}) are the on-lattice version of the angular
dynamics for self-propelled particles considered in \cite{Nagai15}.}

Now we will write an equilibrium-like description of the steady state resulting from the non-equilibrium dynamics
(\ref{motion}). To do so, we employ the Unified Colored Noise approximation (UCNA) \cite{Jung87,Hanggi95} to the many-body problem \cite{Maggi,Marconi1,Marconi2,Marconi3}. 
We start with performing the time derivative
of the first equation in (\ref{motion}). Adopting the dot notation for the time derivative and using the Einstein summation convention,  one has \cite{Maggi}
\BEA \label{inertia}
\tau \ddot{\theta}_i &=& - M_{ij} \dot{\theta}_j - \frac{\p H_{xy}}{\p \theta_i} + \sqrt{T} \eta_i \\ \nn
M_{ij} &\equiv& \delta_{ij} + \tau \frac{\p ^2H_{xy}}{\p \theta_i \p \theta_j} \,,
\EEA
\mttp{ According to (\ref{inertia}), we have rewritten the original set of two first order stochastic differential equations
into a second-order stochastic differential equation where $\tau$ plays the role of inertia
and $M_{ij}$ is the friction. In UCNA one consider the overdamped limit of (\ref{inertia}), to do so
let us introduce $\tilde{M}_{ij}\equiv \tau^{1/2}M_{ij}$ and the rescaled time $z=\tau^{-1/2} t$.
We can then write
\BEQ
\ddot{\theta}_{ij} = -\tilde{M}_{ij} \dot{\theta}_j - \frac{\p H_{xy}}{\p \theta_i} + \tilde{\eta}_i
\EEQ
where for the noise term $\tilde{\eta}$ one has $\langle \tilde{\eta}_i \rangle=0$ and
$\langle \tilde{\eta}_i(z) \tilde{\eta}_j(z^\prime) \rangle=2 T \tau^{-1/2} \delta_{ij} \delta (z - z^\prime)$.
The overdamped limit holds in the large and positive friction limit $\tilde{M}_{ij} \gg 1$.
Since $\tilde{M}_{ij}=\delta_{ij} \tau^{-1/2} + \tau^{1/2} \frac{\p ^2H_{xy}}{\p \theta_i \p \theta_j} $, in the
region of the configuration space where the system is locally stable, i. e., where potential energy hypersurface
has all positive curvatures, the large friction limit is realized in
both situations $\tau\to0$ and $\tau\to\infty$ \cite{Jung87,Hanggi95}. 
In the large friction limit we can write
}
%
\BEA \label{ucna}
\dot{\theta}_i &=& -\frac{1}{2} F_{i}[\theta] + D_{ij}[\theta] \eta_j \\ \nn
F_i[\theta] &\equiv& - 2 M^{-1}_{ij} \frac{\p H_{xy}}{\p \theta_i} \\ \nn
D_{ij}[\theta] &\equiv& \sqrt{T} M^{-1}_{ij} \; .
\EEA
The corresponding Fokker-Planck equation for the probability distribution function $P[\theta,t]$ reads
\BEQ
\p_t P[\theta,t] = \frac{1}{2} \frac{\p}{\p \theta_i} \left\{ 2 D_{ij} \frac{\p}{\p \theta_l}\left[ D_{lj} P \right] + F_i P\right \} \ .
\EEQ
To compute the steady state distribution $P_{ss}[\theta]=\lim_{t \to \infty} P[\theta,t]$, we
consider the solution of $\p_t P[\theta,t] =0$ that is
\BEQ \label{ss}
P_{ss}[\theta]=\det M \exp{ \left( -\frac{H_{xy}[\theta]}{T} - \frac{\tau}{2 T} |\nabla_{\theta_i} H_{xy}|^2 \right) } \cdot Z_{eff}^{-1} \; ,
\EEQ
the numerical constant $Z_{eff}^{-1}$ is the normalization factor.
According to (\ref{ss}), we can write an effective free energy $F_{eff}(T,\tau)$. 
The thermodynamics is then given by the following equations
\BEA\label{effective} 
F_{eff}(N,T,\tau)&=&-T \ln Z_{eff} \\ \nn
Z_{eff} &\equiv& \int_0^{2 \pi} \prod_{i} d\theta_i e^{-\frac{1}{T} H_{eff}[\theta]} \\ \nn
H_{eff}[\theta] &\equiv& H_{xy}[\theta] + \frac{\tau}{2} \left| \nabla_{\theta_i} H_{xy} \right|^2 - T \ln \det M \, .
\EEA 
The presence of $\det M$ in (\ref{effective}) \mttp{makes} the effective 
free energy calculation a hard task that needs further approximations. 
\mttp{As we have discussed before, UCNA holds in the limits $\tau\to0$ and
$\tau\to\infty$. In the first case, even in the presence of negative curvatures, the term $\delta_{ij}$ dominates with respect the Hessian matrix $ \frac{\p^2 H_{xy}}{\p \theta_i \p \theta_j}$.
In that situation the determinant can be computed analytically considering the Hessian
as a small perturbation to the identity matrix. }
\mttp{The mean-field model, i. e., $d=\infty$, will be addressed in Sec. \ref{MFS} considering a fully connected lattice model $J_{ij}=J/N$.
In this way one can compute analytically the partition function in the small $\tau$ limit.
After that, in Sec. \ref{BKTS}, we will study the model for $d=2$ in the continuum limit, i. e., the spin-wave
approximation of $H_{xy}$. In two
dimension, $J_{ij}$ is different from zero only between nearest neighbor sites.} 
%
%

\section{Mean Field approximation and Landau-Ginzburg Free energy} \label{MFS}
\mttp{Here we are interested in investigating the critical properties of $xy-$model, i. e., 
the properties of the system near a second order phase transition. To do so, we
neglect the spatial properties of the system performing the computation (\ref{effective})
on a fully-connected lattice. In this way, we can analytically compute the partition function
and also write the corresponding Landau-Ginzburg theory. The fully connected lattice
is obtained considering an adjacent matrix $J_{ij}=J/N$.}
%
We compute the effective thermodynamics (\ref{effective})
performing a saddle-point approximation to evaluate the partition function $Z_{eff}$ (see Appendix \ref{fully}).
Introducing the free-energy per spin $f(T,\tau)=F_{eff}(N,T,\tau)/N$ and the inverse temperature $\beta=1/T$, one has
\BEA \label{mf2}
f[m] &=&  \frac{\beta J}{2}\left( 1 + \frac{\tau J}{2} \right) m^2 - \ln z[m] \; \\ \nn
z[m] &\equiv& I_0(\beta J m) + \tau m I_1(\beta J m)\, ,
\EEA 
where $m$ is the modulus of the magnetization $\mathbf{m}=(m_x,m_y)=\langle N^{-1} \sum_i \mathbf{s}_i \rangle $. 
We have indicated with $I_n(x)$ the modified
Bessel function of order $n$. 
It is worth noting that (\ref{mf2}) holds only in the small tau limit where we can 
write $\det M\sim 1 + \tau Tr \p_{\theta_i,\theta_j}^2 H_{xy}$.
By minimizing with respect $m$, we obtain the self-consistency equations
\BEQ \label{self}
m - \frac{\left[ 2 I_1(\beta J m) + \tau J \left( I_0(\beta J m) + I_2(\beta J m)\right) \right]}{2 z}  = 0 \, .
\EEQ 
As one can check, when $\tau=0$, 
the equation reduces to the well known mean-field result $m(\beta)=I_1(\beta J m)/ I_0(\beta J m)$. 
\begin{figure}[!t]
\begin{center}
\includegraphics[width=.45\textwidth]{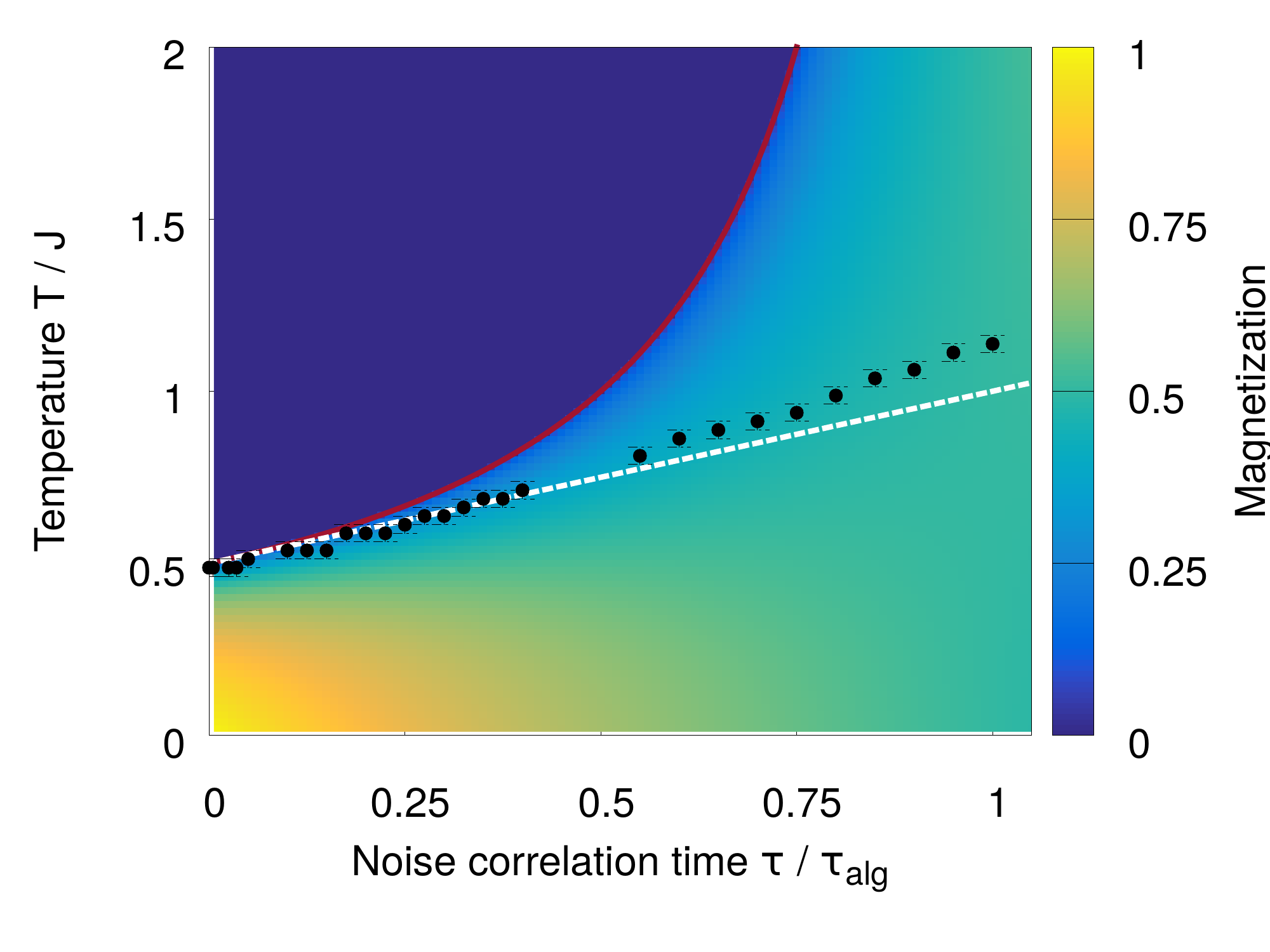}
\caption{Phase diagram. Mean field phase diagram
obtained minimizing the free energy (\ref{mf2}). The red
curve is the analytical computation of the critical line given by (\ref{ct}), black
symbols are numerical simulations, white dashed curve is the small $\tau$
expansion of (\ref{ct}).}
\label{fig:mf}
\end{center}
\end{figure}
%
By expanding (\ref{mf2}) up to the fourth order in $m$ and recalling that 
$m$ is the modulus of the magnetization $\mathbf{m}=(m_x,m_y)$, we obtain the following Landau-Ginzburg free energy
\BEA \label{LG}
f_{LG}[m_x,m_y] &=& \frac{A}{2}(m_x^2 + m_y^2) + \frac{B}{4} (m_x^2 + m_y^2)^2 \\ \nn
A &\equiv& J \beta \left[ 1 - \frac{J}{2} \left( \tau + \beta \right)  \right] \\ \nn
B &\equiv& \frac{J^4 \beta^2}{8} \left[ \frac{\beta^2}{2} + \tau \left( \frac{1}{2} + \tau \beta \right) \right] \; . 
\EEA 
\mttp{$f_{LG}$}
contains all the information we need to understand the critical properties
of the system in mean-field approximation. 
\mttp{For instance, writing $m=|\mathbf{m}|$ and considering the solution 
\BEQ
\frac{\p f_{LG}}{\p m} = 0 \; , \; \frac{\p^2 f_{LG}}{\p m^2} > 0 \; 
\EEQ
we obtain the spontaneous magnetization $m_s$. In the symmetry broken phase one has
$m_s=\sqrt{-A/B}$.
}
According to Eq. (\ref{LG}), the Goldostone picture remains untouched. To realize that we write $\mathbf{m}$
as a complex field parametrized through two real fields $\varphi_1$ and $\varphi_2$, i. e., 
$\mathbf{m} \to \varphi_1 + i \varphi_2$. Looking at the fluctuations near to the minimum of $f_{LG}$, 
in the symmetry broken phase,
we can write $\varphi_1=m_s+\delta \varphi_1$ and $\varphi_2=\delta \varphi_2$. Inserting these two expressions
in $f_{LG}$, one obtains that $m_s^2$ is the mass of the longitudinal fluctuation $\delta \varphi_1$ while the transverse mode $\delta \varphi_2$ is massless, i. e., the Goldstone mode.

To estimate the critical line $T_{mf}(\tau)$ one has to consider the solution of the equation $A=0$ that is
\BEQ \label{ct}
\frac{T_{mf}(\tau)}{J}=\frac{1}{2 - \tau J } \simeq \frac{1}{2} \left[ 1 + \frac{\tau J}{2}  \right] \; .
\EEQ 
\mttp{As we show in (\ref{redundant}), the same critical line can be computed
from the free-energy $f[m]$ given by (\ref{mf2}). 
}
According to (\ref{ct}), we notice that $T_{mf}(\tau)$ increases with $\tau$ and diverges when $\tau=2/J\equiv \tau_{alg}$.
Here, we have introduced a characteristic time scale $\tau_{alg}$ 
that is the time needed to align a spin with the resulting mean-field acting on it.
Since the computation holds at small $\tau$, the divergence is
unphysical. Thus, we have to consider the small $\tau$ expansion of $T_{mf}(\tau)$.

In Fig. (\ref{fig:mf}) we show the resulting phase diagram obtained
by minimizing numerically the free energy (\ref{mf2}). The contour plot represents the magnetization $m(\tau/\tau_{alg},T)$. The red curve is the critical line (\ref{ct}), the black symbols are
obtained by numerical simulations of the fully-connected lattice, 
the details of numerical simulations are in Sec. (\ref{num}).
As one can see, the theoretical prediction reproduces quite well the numerical simulations in the 
small $\tau$ limit, to highlight this finding we have plotted in white the small $\tau$ expansion.
However, deviations from the approximated theory become dramatic by increasing
$\tau$.

\section{Topological Transition in two dimensions} \label{BKTS}

\mttp{In this section, we discuss the effect of persistent noise in $d=2$ where BKT transition takes place at 
temperature $T_{BKT}$ for $\tau=0$.
To do so, we start with considering $H_{xy}$ in the spin-wave approximation \cite{Plischke} that is
}
%
\BEQ\label{sw}
H[\theta(\rr)]=\frac{J}{2 a^{d-2}} \int d^d r \, \nabla \theta(r) \cdot \nabla \theta (r) \, ,
\EEQ
where $a$ is the lattice spacing. 
To write the equation of motion for $\theta(\rr)$, we introduce an auxiliary field $\psi(\rr)$
undergoing an Ornstein-Uhlenbeck process 
\BEA\label{sw2}
\dot{ \theta}(\rr) &=& -\frac{\delta H[\theta]}{\delta \theta(\rr)} + \psi(\rr) \\ \nn
\tau \dot{ \psi}(\rr) &=& -\psi(\rr) + \sqrt{T} \eta(\rr) \, ,
\EEA
the noise term satisfies $\langle \eta(\rr) \rangle=0$ and $\langle \eta(\rr,t) \eta(\rr^\prime,s) \rangle = 2 \delta(\rr - \rr^\prime) \delta(t-s)$. 
\mttp{ By introducing the rescaled time $z=\tau^{-1/2} t$, we can write
\BEA 
\ddot{\theta}(\rr)&=&-\dot{\theta}(\rr)\tilde{M}[\theta(\rr)] - \frac{\delta H}{\delta \theta(\rr)} + \tilde{\eta}(\rr,z) \\ \nn
\tilde{M}[\theta(\rr)] &\equiv&  \frac{1}{\tau^{1/2}} +  \tau^{1/2} \frac{ \delta^2 H}{\delta \theta(\rr) \delta \theta(\rr^\prime)}
\EEA 
where for the noise term $\tilde{\eta}(\rr,z)$ one has $\langle \tilde{\eta}(\rr,z) \rangle = 0$ and
$\langle \tilde{\eta}(\rr,z) \tilde{\eta}(\rr^\prime,z^\prime) \rangle= 2T \tau^{-1/2}\delta(\rr - \rr^\prime)\delta(z - z^\prime)$.
} 
\mttp{In the large friction limit $\tilde{M}\gg1$, we can neglect the inertial term $\ddot{\theta}(\rr)\to 0$.
Since the continuum approximation is performed around the ground state, i. e., where the system
is locally stable, the overdamped dynamics is recovered in the limit $\tau\to 0$ and $\tau\to \infty$.
Here, we will consider the limit $\tau\to 0$, meaning that our results are valid only in the small
$\tau$ limit.
}
\mttp{In that limit and at small enough temperature, the equation of motion for $\theta(\rr)$ reads
\BEQ
\dot{\theta}(\rr) = -\frac{\delta H_{eff}}{\delta \theta(\rr)}
\EEQ
where we have introduced the effective Hamiltonian
\BEA\label{heffsw}
H_{eff}[\theta(\rr)] &=&   H + H_1 \\ \nn
H_1 &\equiv&  \frac{\tau J^2}{2 a^{d-4}} \int d^dr \, \left( \Delta \theta(\rr) \right)^2 \, .
\EEA
}
%
It is convenient to express $\theta(\rr)$ in terms of its Fourier components
$\theta(\rr)=N^{-1/2}\sum_\mathbf{k} \theta_\mathbf{k}e^{i \mathbf{k} \cdot \rr}$.
In this way, we can rewrite the energy as
\BEQ
H_{eff}[\theta] = \frac{a^2 J}{2} \sum_\mathbf{k} k^2 \theta_{\mathbf{k}} \theta_{-\mathbf{k}} + 
\frac{\tau a^4 J^2}{2}\sum_\mathbf{k} k^4 \theta_{\mathbf{k}} \theta_{-\mathbf{k}}
\EEQ
with $k=|\mathbf{k}|$.
As well as the BKT case, we can write $\theta(\rr)=\theta_{sw}(\rr)+\theta_{v}(\rr)$, where $\theta_{sw}$ is the spin wave configuration and $\theta_{v}$ the vortex configuration. 

\begin{figure}[!t]
\begin{center}
\includegraphics[width=.45\textwidth]{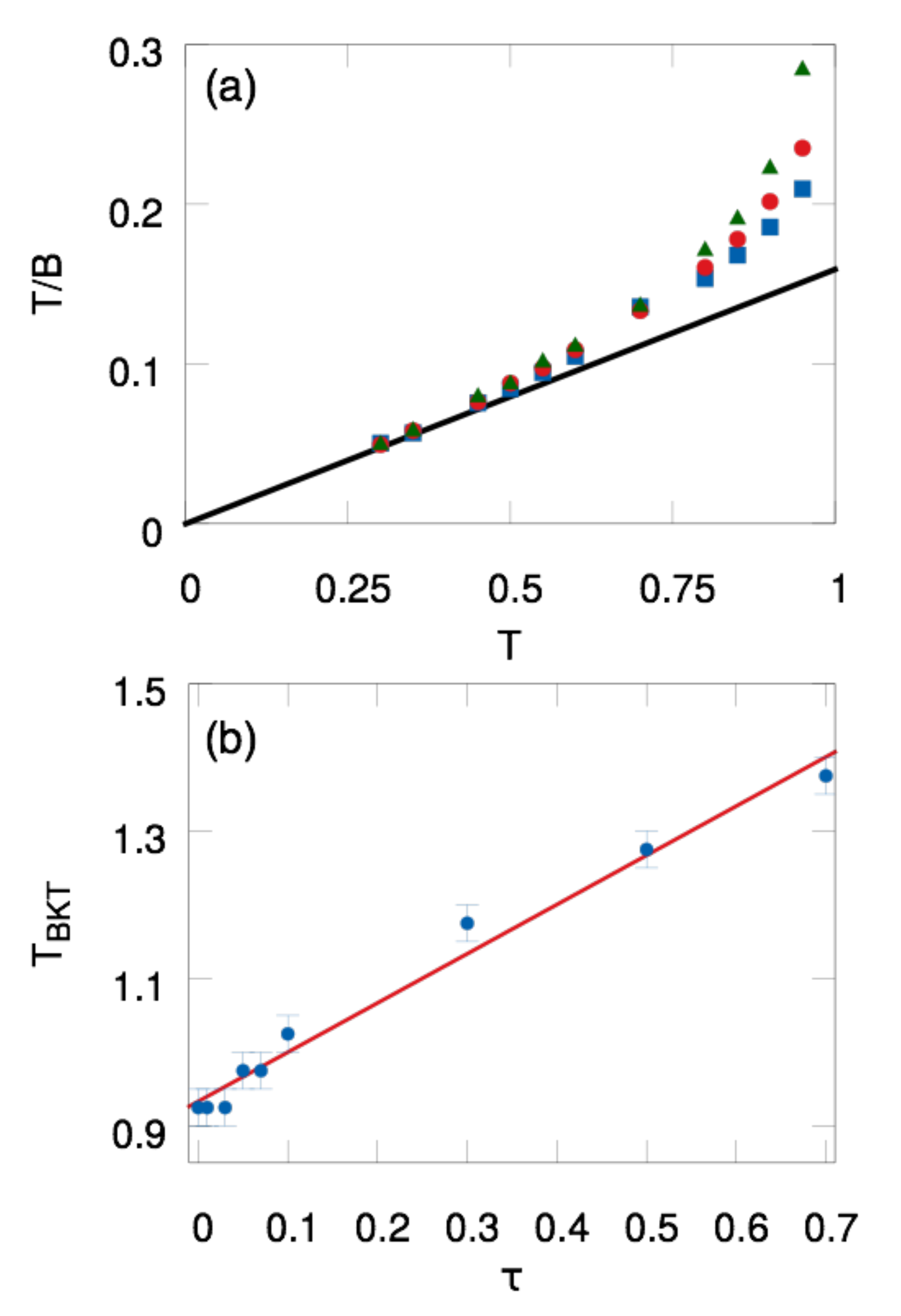}
\caption{Two dimensional simulations. 
(a) Comparison between theory (\ref{corr}) and numerical
simulations. Triangles, circles and squares are $\tau=0.01,0.05,0.1$, respectively. 
At small temperatures $T<0.5$ , data collapse on the same curve. 
(b) Monotonic shift in the temperature of the topological transition 
as a function of $\tau$. \mttp{Blue symbols are simulations, the red line is  
the linear fit $T_{BKT}(\tau)=a+b \tau$, with $a=0.93(1)$ and $b=0.67(3)$.} 
}
\label{fig:2d}
\end{center}
\end{figure}

Now, we can compute the spin-spin correlation function $g(r)=\left\langle e^{-i[\theta_{sw}(\rr) - \theta_{sw}(0)]}\right\rangle$, the details are \mttp{discussed} in Appendix \ref{corre}. In the limit $r^2 \gg a^2 J \tau $ one has
\BEQ\label{corr}
g(r)\sim \left( \frac{\pi r}{a} \right)^{-\frac{T}{2 \pi J}} \left( 1 + J \pi \tau \right)^{\frac{T}{4 \pi J}}.
\EEQ 
Because $g(r\to\infty)=0$, Eq. (\ref{corr}) implies also that, as well the equilibrium case, no long-range
order is found for an infinite system. However, for a finite-size system, if $r$ is of
the order of the system size and $a^2 J \tau \gg r^2$, we have that 
$g(r)\simeq 1$, i. e., the system is practically in the ground state with all the spins aligned.
In other words, \mttp{at low enough temperatures}, memory in the noise promotes uniform
configuration suppressing long wave length excitations
, at least in the small $\tau$ regime. 
\mttp{
To check the validity of that prediction, we have computed numerically the spin-spin correlation
function $g(r)$, the details of the simulations are given in Appendix \ref{num}.
In particular, we have fitted the numerical data 
to the functional form 
$g_{fit}(r)=(A r)^{-T/B}(1+\frac{B \tau}{2})^{T/2B}$.
In Fig. (\ref{fig:2d}-a) we show the behavior of $T/B$ vs $T$ for $\tau=0.01,0.05,0.1$, squares, circles, and triangles, respectively. 
The black line is the theoretical prediction (\ref{corr}), i. e.,  $B=2 \pi J$.
As one can see, for temperatures $T<0.5$, the data collapse on (\ref{corr}).
}

\mttp{ Now we estimate the energy cost of a single vortex in the presence
of correlated noise. In this way, we can quantify the effect of persistent noise
on the BKT temperature. The vortex configurations $\theta_v(\rr)$ minimize
(\ref{heffsw}) and satisfy the boundary condition}
$
\oint d\mathbf{l} \cdot \nabla \theta_{v} = 2 n  \pi \, .
$
\mttp{As well in the equilibrium case, also in the small $\tau$ limit, vortices have
the form}
$\nabla_r \theta_{v}(r)=1/r$. 
Inserting the vortex configuration in $H_{eff}$, we can compute
the free energy cost $\Delta f_{vortex}=H-k_B T S$, where $S$ is the entropy of a single vortex. 
Performing a straightforward calculation, one can obtain
$\Delta f_{vortex}$ that is
\BEA \label{deltaf}
\Delta f_{vortex} &=& H_0 + H_1 - 2 k_B T \ln \frac{L}{a} \\ \nn 
&=& \left[ \pi J - 2 k_B T \right] \ln \frac{L}{a} \\ \nn
&+& \frac{\pi \tau J^2}{2} \left[ 1 - \frac{a^2}{L^2} \right] \, .
\EEA
According to (\ref{deltaf}), the exponentially correlated noise produces a shift $\Delta T_{BKT}$ 
in the critical temperature of the topological transition that is
\BEA \label{topo}
\Delta T_{BKT}(\tau) &=& \frac{\pi J^2 g(L,a)}{4 \ln{(L/a)} } \tau \\ \nn
g(L,a) &\equiv&  1 - \frac{a^2}{L^2} \, .
\EEA
As one can see, $\Delta T_{BKT}$ is linear in $\tau$. However, since the linear size of the system $L$ grows with $N^{1/2}$,
in the thermodynamic limit $\lim_{N\to\infty} \Delta T_{BKT} = 0$, meaning that the location
of the topological transition remains untouched. 
Considering a finite size system, we can define a size-dependent crossover temperature 
\mttp{$T_{BKT}(\tau,N)$. 
According to (\ref{topo}), one has $\lim_{N\to\infty}T_{BKT}(\tau,N)=T_{BKT}$. However,
at finite $N$,} we expect to observe 
a linear shift in $\tau$ towards higher temperature.
\mttp{
We have tested that prediction in numerical simulations.
The resulting  shift in $T_{BKT}$ is shown in Fig. (\ref{fig:2d}-b). As one can see, it is in good agreement with (\ref{topo}). The blue line is obtained fitting the data to $T_{BKT}(N,\tau)=a + b \tau$, where $a$ and $b$ are the fitting parameters. The parameter $a \equiv T_{BKT} (0)\sim 0.9$ is compatible with recent accurate estimate of $T_{BKT}$ \cite{Berganza13,Hasenbusch05}.}
%


\mttp{
\section{Summary and Discussion}
In this paper, we have proposed an effective equilibrium theory
for the $xy-$model driven out-of-equilibrium by exponentially 
correlated noise. To map the original many-body problem
into an effective equilibrium picture, we have employed 
UCNA \cite{Jung87,Hanggi95}. 
In the results presented here, the persistent time $\tau$ plays the role
of external and tunable thermodynamic parameter. 
Moreover, even though UCNA should work also in the $\tau\to\infty$
limit, in the many-body case the presence of $\det M$ requires, in
general, further approximations \cite{Maggi}. Since the matrix $M$ has
the form $M=\mathbb{1}+\tau \mathbb{H}$, where $\mathbb{1}$ is the
identity matrix and $\mathbb{H}$ the Hessian matrix, we have considered
the approximation $\det M = 1 + \tau Tr \mathbb{H}$, that holds in the small $\tau$
limit.
%
%
We have specialized our computation in two cases: (i) $d=\infty$ corresponding to 
the mean-field approximation, and (ii) $d=2$, where, in equilibrium, BKT transition takes place.  
}
%
\mttp{Differently from the scalar field case, where a Landau $\varphi^4$ theory has been 
proposed phenomenologically to describe the impact of correlated
noise on critical phenomena in Active Matter\cite{Paoluzzi}, the mean-field computation
presented here allows to obtain the coarse-grained theory starting
from a microscopical model. 
In particular}, we have computed analytically the effective partition function and, 
expanding the free energy 
\mttp{around} the transition point, we have 
\mttp{obtained} the corresponding Landau-Ginzburg free energy \mttp{$f_{LG}$}. 

\mttp{We have shown that
the coefficient of the quadratic term of \mttp{$f_{LG}$} vanishes along
the critical line $T_{mf}(\tau)$. Moreover, the resulting $T_{mf}(\tau)$  is an
increasing function of $\tau$, i. e., starting from a disorder configuration at high $T$
and maintaining $T$ fixed, the memory of the noise can be tuned to
bring the system at criticality. This property of non-equilibrium models driven by
exponentially correlated noise seems to be quite general since it has been already observed 
in both, theory and numerical simulations in the case
of zero-dimensional $\varphi^4$ theory with exponentially correlated noise \cite{Paoluzzi}, i. e., 
the gas-liquid universality class, and also in the case of the glassy transition of active
particles driven by colored noise \cite{Szamel15,Nandi16,Nandi17}.} 

\mttp{To check the validity of that finding we have 
performed numerical simulations of the fully-connected model. The critical points
in the small $\tau$ regime obtained from numerical simulations 
follow quite well the theoretical prediction $T_{mf}(\tau)$.}
\mttp{Since $f_{LG}$ describes an $O(2)$ vectorial field theory, 
crossing the critical line the symmetry $O(2)$ is spontaneously broken and,
according to the Goldstone mechanism \cite{Goldstone}, the longitudinal fluctuations
are massless, while the mass of the transverse excitation depends on $\tau$.} 


After that, we have studied the theory in two dimensions where 
BKT transition takes place. We have shown that, in the \mttp{small $\tau$ limit}, 
the topological transition remains untouched by the non-equilibrium dynamics.
However, considering a finite size system, 
the theory predicts a linear shift in $T_{BKT}$ meaning that memory
disadvantages vortex excitations. Thus, at low temperatures, 
the non-equilibrium system turns to be more correlated than 
the equilibrium counterpart. Performing numerical simulations
in two dimensions, we found a good qualitative agreement between
theory and numerics.

\mttp{
It would be very interesting to try to extend these approximation schemes 
to off-lattice models.
In this way, one could estimate the impact of memory
effects on the collective properties of
assemblies of self-propelled particles with alignment interactions \cite{Vicsek12}}.
\mttp{Recently, it has been shown in both experiments and models, 
that memory effects in the angular dynamics play an important role \cite{Nagai15,Sumino12}. It is wort noting that 
the well established methods describing collective properties of self-propelled
particles can not be applied in the case of exponentially correlated dynamics \cite{Marchetti13}.
For instance, analytical predictions about the effects of exponentially correlated noise
on angular dynamics can be made only in the low-density limit and
considering a simplified one-dimensional telegraphic noise model for
describing the memory effects \cite{Nagai15}. According to our computation scheme,
in the small $\tau$ limit, memory effects in the angular dynamics can be reabsorbed into an effective
equilibrium Hamiltonian $H_{eff}=H_{xy}+\frac{J \tau}{2} | \nabla_\theta H_{xy}|^2$.
%
%
}

\section*{Acknowledgments}
We thank M. Cristina Marchetti for illuminating discussions and S. Rold\'an Vargas
for his critical reading of the manuscript, 
MP was supported by the Simons Foundation  
Targeted Grant in the Mathematical Modeling of Living Systems Number: 342354 and by the Syracuse Soft Matter Program.
C. Maggi acknowledges support from the European Research Council under the European Union's Seventh Framework programme
(FP7/2007-2013)/ERC Grant agreement  No. 307940.

\appendix
\section{Fully connected model} \label{fully}
The mean-field solution of the $xy-$model has been computed
considering a fully-connected lattice \mttp{that corresponds to $d=\infty$ situation \cite{ZinnJustin,Parisi_glass,Parisi_stat}.} 
The Hamiltonian reads
\BEQ 
H_{xy}^{MF}[\theta] = -\frac{J}{2 N} \sum_{i,j}\cos(\theta_i - \theta_j) \, .
\EEQ 
To compute the partition function we introduce the following order parameters
\BEA  
N \phi &=& \sum_i \cos \theta_i, \;\; N \psi = \sum_i \sin \theta_i  \\ \nn
N \pi &=& \sum_i \cos 2 \theta_i, \;\; N \sigma = \sum_i \sin 2 \theta_i  \, ,
\EEA  
in terms if the order parameters the Hamiltonian can be written as follows
\begin{multline}
H_{xy}^{MF}[\theta] = -\frac{N J}{2} \left( 1 - \frac{\tau J}{2} \right) \left[ \phi^2 + \psi^2 \right] + \\ 
+ \frac{N \tau J^2}{4} \left( \psi^2 - \phi^2 \right) \pi  + \\
- \frac{N \tau J^2}{2} \phi \psi \sigma - \frac{1}{\beta} \ln \det M \, .
\end{multline} 
In the small $\tau$ limit, we approximate the determinant in the following way
\BEQ 
\det M \simeq 1 + \tau Tr \frac{\p^2 H_{xy}^{MF}}{\p \theta_i \p \theta_j}
\EEQ 
and the trace of the Hessian matrix reads
\BEQ 
Tr \frac{\p^2 H_{xy}^{MF}}{\p \theta_i \p \theta_j} = J \sum_i \left( \phi \cos \theta_i + \psi \sin \theta_i \right) + \mathcal{O}(\frac{1}{N}) \; .
\EEQ 
To compute the partition function we represent the order parameters $(\phi,\psi,\pi,\sigma)$ 
through a set of lagrangian multipliers $\lambda_k$, with $k=1,...,4$, as follows
\BEA \nn
\delta\left( N \phi - \sum_i \cos \theta_i \right)   &=&  \int d\lambda_1 \, e^{ -\lambda_1 \left( N \phi       - \sum_i \cos \theta_i \right)} \\ \nn
\delta\left( N \psi       - \sum_i \sin \theta_i \right)    &=&  \int d\lambda_2\, e^{ -\lambda_1\left( N \psi       - \sum_i \sin \theta_i \right)} \\ \nn
\delta\left( N \pi         - \sum_i \cos 2\theta_i \right) &=&  \int d\lambda_3\, e^{ -\lambda_1 \left( N \pi         - \sum_i \cos 2\theta_i \right)} \\ \nn
\delta\left( N \sigma - \sum_i \sin 2\theta_i \right)  &=&  \int d\lambda_4\, e^{ -\lambda_1 \left( N \sigma - \sum_i \sin 2\theta_i \right)}
\EEA
finally, the partition function reads
\BEA
Z&=&\mathcal{N} \int d\Phi\, e^{-N f}, \,\, \Phi\equiv(\lambda_i,\phi,\psi,\pi,\sigma) \\ \nn
f&\equiv& -\frac{\beta J}{2} \left( 1 - \frac{\tau J}{2}\right) \left( \phi^2 + \psi^2 \right) + \\ \nn 
&+&\frac{\beta J^2 \tau}{4} \left( \psi^2 - \phi^2\right) \pi - \frac{1}{2}\beta J^2 \tau \phi \psi \sigma + \\ \nn
&+&\lambda_1 \phi + \lambda_2 \psi + \lambda_3 \pi + \lambda_4 \sigma - \log z \\ \nn
z&\equiv&\int_0^{2\pi}d\theta\,A(\phi,\psi)_{\theta}\, e^{-\mathcal{H}^\prime} \\ \nn
A(\phi,\psi)_{\theta}&\equiv&1 + \tau J (\phi \cos{\theta} + \psi \sin{\theta}) \\ \nn
-\mathcal{H}^\prime&\equiv& \lambda_1 \cos{\theta} + \lambda_2 \sin{\theta} + \lambda_3 \cos{2 \theta} + \lambda_4 \sin{2 \theta}
\EEA
where $\mathcal{N}$ is a normalization constant.   
\subsection{Saddle-point equations}
In the thermodynamic limit $N\to\infty$, we can perform the saddle-point approximation to evaluate
the partition function \cite{ZinnJustin,Parisi_stat,Parisi_glass}
\BEQ
Z\sim e^{-N f_{SP}}, \;\; \left. \frac{\p f}{\p \Phi}\right|_{SP} = 0
\EEQ 
and the saddle-point equations are
\mttp{
\BEA \nn
\lambda_1&=&  \frac{\beta J^2 \tau}{2} (\phi \pi + \psi \sigma) + \beta J (1 - \frac{\tau J}{2}) \phi + I_1 \\ \nn
\lambda_2&=&  \frac{\beta J^2 \tau}{2} (\phi \sigma - \psi \pi)  + \beta J (1 - \frac{\tau J}{2}) \psi  + I_2 \\ \nn
\lambda_3&=&  \frac{\beta J^2 \tau}{4}(\phi^2 - \psi^2) \\ \nn
\lambda_4&=&  \frac{1}{2}\beta J^2 \tau \phi \psi \\ \nn
\phi      &=& \langle \cos{\theta} \rangle_{\mathcal{H}}\;, \;\; \psi      = \langle \sin{\theta} \rangle_{\mathcal{H}} \\ \nn
\pi        &=& \langle \cos{2 \theta} \rangle_{\mathcal{H}}\;, \;\; \sigma = \langle \sin{2 \theta} \rangle_{\mathcal{H}} \\ \nn
-\mathcal{H} &\equiv& -\mathcal{H}^\prime + \log{A(\phi,\psi)_{\theta}} \\ \nn
I_1 &\equiv& \frac{\tau J}{z} \int d\theta \, \cos \theta \, e^{-\mathcal{H}^{\prime}} \\ 
I_2 &\equiv& \frac{\tau J}{z} \int d\theta \, \sin \theta \, e^{-\mathcal{H}^{\prime}} 
\EEA
where we have introduced the average of a generic observable with respect the effective one-body Hamiltonian $\mathcal{H}$ that is $\langle \mathcal{O} \rangle_{\mathcal{H}} \equiv \frac{\int d\theta \mathcal{O} e^{-\mathcal{H}}}{\int d\theta\, e^{-\mathcal{H}}}$.}

\subsection{Elimination of the redundant variables} \label{redundant}
In order to eliminate the redundant variables that we have introduced to compute the partition function, we
write the auxiliary fields in polar coordinates
\BEA
\phi &=& m \cos \Theta \,, \;\; \lambda_1 = \Lambda \cos \lambda \,,\;\;  \lambda_3 = n \cos \hat{\lambda} \\ \nn
\psi &=& m \sin \Theta  \,,  \;\; \lambda_2 = \Lambda \sin \lambda \,,\;\;  \lambda_4 = n \sin \hat{\lambda}
\EEA
From the equations for $\lambda_{3,4}$ it follows that $n=\hat{\lambda}=0$, and, as a consequence,
$\pi=\sigma=0$.
The free energy of the model can be written as follows
\BEA \label{fe} \nn
f[m,\Lambda] &=& -\frac{\beta J}{2} \left( 1 - \frac{\tau J}{2} \right) m^2 + m \Lambda - \log z \\ 
z &=& I_0(\Lambda) + \tau J m I_1(\Lambda) \, .
\EEA
The self-consistency equations are
\BEA \label{sc} 
\frac{\p f}{\p m} &=& -\beta J (1 - \frac{\tau J}{2} )m + \Lambda - \frac{\tau J I_1(\Lambda)}{z}= 0 \\ \nn
\frac{\p f}{\p \Lambda} &=& m - \frac{1}{2 z} \left[ 2 I_1(\Lambda) + \tau J \left( I_0(\Lambda) + I_2(\Lambda) \right) \right] = 0 \, .
\EEA
When $\tau=0$ we recover the mean field solution of the equilibrium $xy$ model
\BEA
\Lambda &=& \beta J m \\ \nn
m &=& \frac{I_1(\beta J m)}{I_0(\beta J m)}
\EEA
Moreover, from (\ref{sc}) one has $\Lambda = \beta J m + \mathcal{O}(\tau^2)$.
Plugging this relation in (\ref{fe}) we obtain (\ref{mf2}).
The critical line $T_{mf}(\tau)$ can be computed considering the solution of 
$\left. \p^2_m f[m] \right|_{m=0}=0$. The computation \mttp{brings to} the same result obtained in the main
text (\ref{ct}) that has been obtained considering the Landau-Ginzburg free energy (\ref{LG}). 
%

\section{Spin waves} \label{sw}
At low enough temperature the $xy$ Hamiltonian in $d$ spatial dimensions can be written as follows
\BEQ
H=\frac{J}{2 a^{d-2}} \int d^d r \, \nabla \theta(r) \cdot \nabla \theta (r) \, .
\EEQ
As we have shown in the main text, the effective Hamiltonian in the small noise limit reads
\BEQ
H_{eff}\equiv H + \frac{\tau}{2} \left| \frac{\delta H}{\delta \theta} \right|^2 \, .
\EEQ
Now, to evaluate the second term on the left-hand side of the last equation, we come back
to the lattice model. \mttp{When $d<\infty$, $J_{i,j}=J\neq0$ only for nearest neighbor sites $i,j$
that we indicate $<i,j>$. The Hamiltonian reads}
\BEQ
H=-J \sum_{<i,j>}\cos(\theta_i - \theta_j) \simeq -J\sum_{<i,j>} \left[  1 - \frac{(\theta_i - \theta_j)^2}{2} \right]
\EEQ
the derivative with respect $\theta_i$ reads
\BEQ
\frac{\p H}{\p \theta_i} = - J \sum_{i} \left( \theta_{i+a}  + \theta_{i-a} - 2 \theta_i \right)= -J a^2 \Delta \theta_i
\EEQ
where $\Delta$ is the Laplace operator. In the continuum limit the Hamiltonian becomes
\BEA 
H_{eff} &=&   H + H_1 \\ \nn
H_1 &\equiv&  \frac{\tau J^2}{2 a^{d-4}} \int d^dr \, \left( \Delta \theta(\rr) \right)^2 \, .
\EEA
\section{Spin-spin correlation function}\label{corre}
Now we compute the spin-spin correlation function $g(r)$ that is \cite{Plischke}
\BEQ
g(r)=\exp{ \left\{  -\frac{1}{2} \int \prod_\kk d \theta_\kk e^{-\frac{H_{eff}}{T} }  \left[ \theta(\rr) - \theta(0) \right]^2 \right\} }
\EEQ
which \mttp{is} 
\BEQ \label{gofr}
g(r)=\frac{T}{N} \sum_\kk \frac{1 - \cos \left( \kk \cdot \rr \right) }{a^2 J k^2 + a^4 J k^4 \tau} \, .
\EEQ 
Now we switch to the continuum also in $k-$space by setting 
$N^{-1}\sum_\kk... \to \frac{a^d}{(2 \pi)^d} \int d\kk ...$ so that Eq. (\ref{gofr}) becomes
\BEQ \label{gofr2}
g(r) = T \left( \frac{a}{2 \pi} \right)^d \int d\kk\, \frac{1 - \cos \left( \kk \cdot \rr \right) }{a^2 J k^2 + a^4 J k^4 \tau} \, .
\EEQ
Specializing the calculation to $d=2$ case, by integrating Eq. (\ref{gofr2})
in polar coordinates, introducing the Bessel function $J_0(x)$, we get
\BEQ \label{gofr3}
g(r) = T \left( \frac{a}{2 \pi} \right)^2 \int dk\, 2 \pi k \frac{1 - J_0(kr)}{a^2 J k^2 + a^4 J k^4 \tau} \, .
\EEQ
When $r$ is very large compared with $a$, we can neglect the Bessel function and 
approximate Eq. (\ref{gofr3}) as
\BEQ
g(r) \simeq \frac{\pi r}{a}^{-\frac{T}{2 \pi J}} \left[ \frac{r^2 + a^2 J \tau }{r^2 \left( 1 + J \pi \tau\right) } \right]^{-\frac{T}{4 \pi J}}
\EEQ
when $r^2 \gg a^2 J \tau$, we recover the result Eq. (\ref{corr}) of the main text.

\mttp{
\section{Numerical simulations} \label{num}
We have solved numerically the equations of motion (\ref{motion})
where the $N$ compasses $\mathbf{s}_{\rr}=(\cos\theta_\rr,\sin\theta_\rr)$ 
are arranged on a two dimensional square lattice. The
vector $\rr$ with $\rr = i \mathbf{x} + j \mathbf{y}$ individuates
the site $(i,j)$ of the lattice, with $i,j=1,..,\sqrt{N}$.
The connectivity of the adjacent matrix $J_{ij}$ defines the spatial
dimensions $d$ where the model is embedded. 
In finite dimensions $d$, $J_{ij}=J$ among nearest neighbor sites.  
The mean-field consists in a fully-connected lattice, i. e., $J_{ij}=J/N$, $\forall i,j$.
Here, we report the results
concerning $N=4900$ ($d=2$) and $N=400$ (fully-connected lattice). The equations of motion are integrated
using a second-order Runge-Kutta scheme with integration
time step $dt=10^{-3}$.
\subsection{Two dimensions}
In two dimensions, we have computed the correlation function
\BEQ \label{num_corr}
g(\rr)=\left\langle N^{-1} \sum_{\rr^\prime} \mathbf{s}_{\rr + \rr^\prime  } \cdot \mathbf{s}_{\rr^\prime} \right\rangle_{t} \, ,
\EEQ
where the average $\langle \cdot \rangle_t$ in (\ref{num_corr}) of a generic 
observable $\mathcal{O}[\theta(t)]$ is computed averaging over one long trajectory of the system 
with a single noise realization, i. e., $\langle \mathcal{O} \rangle_t=t^{-1}\int_{t_0}^{t_0 + t}ds\, \mathcal{O}[\theta(s)]$.
}

\mttp{
The transition temperature $T_{BKT}(\tau)$ has been computed considering a power
law fit to $r^{-\eta}$ for the spatial correlation function $g(r)$. We define $T_{BKT}(\tau)$
using the criterium $\eta=\frac{1}{4}$ at the transition temperature \cite{PRBeta}.
%
\subsection{Mean Field}
In Fig. (\ref{fig:mf}), we compare the mean-field prediction (\ref{ct}) with numerical simulations
of a fully-connected lattice composed by $N=400$. We have considered $30$ temperatures for each $\tau$ with
$\tau\in[10^{-3},2]$.
The critical point has been obtained considering the modulus of the magnetization $m=\sqrt{m_x^2+m_y^2}$, where
$m_{x}=N^{-1}\sum_\mathbf{r} \cos \theta_{\mathbf{r}}$, and $m_{y}=N^{-1}\sum_\mathbf{r} \sin \theta_{\mathbf{r}}$ are the magnetization along $x$ and $y$, respectively.
To evaluate the transition temperature, we have looked at the susceptibility $\chi=N \langle \left( m - \langle m \rangle\right)^2 \rangle $ that develops a peak at the transition.
}

\bibliography{mpbib}
\end{document}